\documentclass[12pt]{article}
\usepackage{amsmath,amssymb}
\usepackage{bm}
\usepackage{mathrsfs}
\usepackage{fourier}
\usepackage{multirow}
\allowdisplaybreaks[4]
\newcommand{{\SlashD}}{D\!\!\!\!\!\!\big/}
\newcommand{{\Slashq}}{q\!\!\!\!\!\big/}
\newcommand{{\SlashF}}{{\rm F}\!\!\!\!\!/}
\usepackage[usenames]{color}

\begin{document}

\title{Chasing after flavor symmetries 
of quarks from bottom up}

\author{
Yoshiharu \textsc{Kawamura}\footnote{E-mail: haru@azusa.shinshu-u.ac.jp}\\
{\it Department of Physics, Shinshu University, }\\
{\it Matsumoto 390-8621, Japan}\\
}

\date{
April 2, 2019}

\maketitle
\begin{abstract}
We explore a flavor structure of quarks in the standard model
under the assumption that flavor symmetries exist 
in a theory beyond the standard model,
and chase after their properties, using a bottom-up approach.
We reacknowledge that a flavor-symmetric part of Yukawa coupling matrix
can be realized by a rank-one matrix 
and a democratic-type one occupies a special position,
based on Dirac's naturalness.
\end{abstract}

\section{Introduction}

The Yukawa sector in the standard model (SM) holds many mysteries.
For instance, the origin of the fermion mass hierarchy and flavor mixing is a big riddle.
There have been many intriguing attempts to explain the values
of physical parameters concerning the fermion masses and
flavor mixing matrices, based on the top-down approach~\cite{CEG,F,HHW,F2,GJ,FN},
but we have not arrived at a satisfactory answer.

There are several reasons why it is difficult to understand 
an origin of the flavor structure.
First, we have no powerful guiding principle 
to determine a theory beyond the SM.
Although flavor symmetries are possible candidates\footnote{
The flavor structure of quarks and leptons has been studied
intensively, based on various flavor symmetries~\cite{FN,MY,I,HPS,HS,AF,IKOSOT,IKOOST}.
},
any evidence has not yet been discovered.
If they exist at all, they might be hidden 
in a false bottom of the Yukawa interactions.
In concrete, there are no unbroken flavor-dependent symmetries in the SM~\cite{LNS,Koide}.
There can be several existence forms
of flavor symmetries in a broken phase of an underlying theory.
For instance, flavor symmetries are broken down in every interactions,
they (or those sub-symmetries) survive in some interactions,
or a new symmetry appears in some terms.
Except for the first one, 
Yukawa interactions, in general, consist of flavor-symmetric and breaking parts
and they are not reconstructed
from experimental data alone 
because global U(3) symmetries emerge in the fermion kinetic terms of the SM.
In other words, there is no way to determine 
the fermion masses and mixing angles without any excellent new concept.
Furthermore, fermions in the SM do not necessarily behave as
unitary bases of flavor symmetries, i.e.,
quarks and leptons are transformed 
using elements of a flavor group G$_{\rm F}$ realized by non-unitary matrices
if an underlying theory possesses non-canonical matter kinetic terms~\cite{YK}.

The world of flavor can be glimpsed from
the Lagrangian in the SM by adopting Dirac's naturalness.
Here, Dirac's naturalness means
that {\it the magnitude of dimensionless parameters on terms allowed by symmetries
should be $O(1)$ in a fundamental theory}
and suggests that the Yukawa coupling of top quark can originate from
a flavor-symmetric renormalizable interaction.
In contrast, other tiny Yukawa couplings are expected to 
come from non-renormalizable ones
suppressed by a power of a high-energy scale.
Then, we obtain a conjecture that {\it a flavor-symmetric part of 
up-type quark Yukawa coupling matrix
can be realized by a rank-one matrix
and a democratic-type one can take a peculiar position}.

If flavor symmetries exist in an underlying theory
and the flavor structure in the SM appears after the breakdown of G$_{\rm F}$,
it is desirable to study the above conjecture 
using suitable field variables such as unitary bases of G$_{\rm F}$.
Although a same conclusion ought to be obtained
because of no change of physics by a choice of field variables,
there is a possibility that it provides a clue to figure out 
the origin of flavor and it gives us a new insight of flavor physics.

In this paper, we explore the flavor structure a little further,
adopting Dirac's naturalness,
and re-examine whether the above conjecture holds or not.

The outline of this paper is as follows.
In the next section, we explain our setup on Yukawa interactions of quarks.
In Sect. 3, we chase after properties of flavor symmetries.
In the last section, we give conclusions and discussions.

\section{Setup}

We explain the setup of our analysis~\cite{YK}.
Our basic assumptions are as follows.
(a) There are flavor symmetries beyond the SM.
(b) The symmetries are broken down by the vacuum expectation values (VEVs) of flavons,
on the whole.
Some symmetries can survive or emerge in some terms.
(c) Flavons also couple to matter fields through matter kinetic terms.

Let us start with a theory of quarks beyond the SM, 
described by the Lagrangian density:
\begin{eqnarray}
&~& \mathscr{L}_{\rm BSM}^{\rm quark} 
= K_{ij}^{(q)} \overline{q}'_{{\rm L}i} i \SlashD q'_{{\rm L}j}
+ K_{ij}^{(u)} \overline{u}'_{{\rm R}i} i \SlashD u'_{{\rm R}j}
+ K_{ij}^{(d)} \overline{d}'_{{\rm R}i} i \SlashD d'_{{\rm R}j}
\nonumber \\
&~&  ~~~~~~~~~~~~~~~~~~ - \left(Y_1\right)_{ij} \overline{q}'_{{\rm L}i} \tilde{\phi} u'_{{\rm R}j}
- \left(Y_2\right)_{ij} \overline{q}'_{{\rm L}i} \phi d'_{{\rm R}j} + {\rm h.c.},
\label{L-BSM-quark}
\end{eqnarray}
where $q'_{{\rm L}i}$ are counterparts of left-handed quark doublets,
$u'_{{\rm R}i}$ and $d'_{{\rm R}i}$ are those of
right-handed up- and down-type quark singlets,
$i, j (=1, 2, 3)$ are family labels, 
summation over repeated indices is understood,
$\phi$ is the Higgs doublet, $\tilde{\phi} = i \tau_2 \phi^*$,
and h.c. stands for the Hermitian conjugation of former terms.
The $K_{ij}^{(q)}$, $K_{ij}^{(u)}$, $K_{ij}^{(d)}$, 
$\left(Y_1\right)_{ij}$, and $\left(Y_2\right)_{ij}$
contain flavons such that
$\mathscr{L}_{\rm BSM}^{\rm quark}$ is invariant under
transformations relating to flavor symmetries.
The $q'_{\rm L}$, $u'_{\rm R}$, and $d'_{\rm R}$ are 
unitary bases of a flavor group ${\rm G}_{\rm F}$,
and are transformed as
\begin{eqnarray}
q'_{\rm L} \to F_{\rm L} q'_{\rm L},~~
u'_{\rm R} \to F_{\rm R}^{(u)} u'_{\rm R},~~d'_{\rm R} \to F_{\rm R}^{(d)} d'_{\rm R},~~
\phi \to e^{i \theta} \phi,
\label{F-tr-prime}
\end{eqnarray}
where $F_{\rm L}$, $F_{\rm R}^{(u)}$, and $F_{\rm R}^{(d)}$ 
are $3 \times 3$ unitary matrices which are elements of ${\rm G}_{\rm F}$
and family labels are omitted.
From the G$_{\rm F}$ invariance of $\mathscr{L}_{\rm BSM}^{\rm quark}$,
we obtain relations:
\begin{eqnarray}
&~& F_{\rm L} K^{(q)} F_{\rm L}^{\dagger} = K^{(q)},~~
F_{\rm R}^{(u)} K^{(u)} F_{\rm R}^{(u)\dagger} = K^{(u)},~~
F_{\rm R}^{(d)} K^{(d)} F_{\rm R}^{(d)\dagger} = K^{(d)},~~
\label{F-tr-K}\\
&~& e^{i\theta} F_{\rm L} Y_1 F_{\rm R}^{(u)\dagger} = Y_1,~~
e^{-i\theta} F_{\rm L} Y_2 F_{\rm R}^{(d)\dagger} = Y_2.
\label{F-tr-Y}
\end{eqnarray}
The $\mathscr{L}_{\rm BSM}^{\rm quark}$
describes only the part relating to quarks in new physics,
and chiral anomalies are supposed to be canceled by other contributions
if the G$_{\rm F}$ symmetries are local.

We assume that G$_{\rm F}$ changes into H$_{\rm F}^{\rm k}$ and H$_{\rm F}^{\rm y}$
after flavons acquire the VEVs at some high-energy scale $M_{\rm BSM}$.
Here, H$_{\rm F}^{\rm k}$ and H$_{\rm F}^{\rm y}$ are
flavor groups of quark kinetic terms and Yukawa interactions, respectively.
Then, $\mathscr{L}_{\rm BSM}^{\rm quark}$ turns out to be
the Lagrangian density: 
\begin{eqnarray}
&~& {\mathscr{L}'}_{\rm SM}^{\rm quark} 
= k_{ij}^{(q)} \overline{q}'_{{\rm L}i} i \SlashD q'_{{\rm L}j}
+ k_{ij}^{(u)} \overline{u}'_{{\rm R}i} i \SlashD u'_{{\rm R}j}
+ k_{ij}^{(d)} \overline{d}'_{{\rm R}i} i \SlashD d'_{{\rm R}j}
\nonumber \\
&~& ~~~~~~~~~~~~~~~~~~~~ - \left(y_1\right)_{ij} \overline{q}'_{{\rm L}i} \tilde{\phi} u'_{{\rm R}j}
- \left(y_2\right)_{ij} \overline{q}'_{{\rm L}i} \phi d'_{{\rm R}j} + {\rm h.c.},
\label{L-SM-quark-prime}
\end{eqnarray}
where $k_{ij}^{(q)}$, $k_{ij}^{(u)}$, and $k_{ij}^{(d)}$ are quark kinetic coefficients,
and $\left(y_1\right)_{ij}$ and $\left(y_2\right)_{ij}$ are 
Yukawa couplings in the unitary bases of G$_{\rm F}$.
Note that non-canonical matter kinetic terms 
appear in $ {\mathscr{L}'}_{\rm SM}^{\rm quark}$.\footnote{
Several works on the flavor physics have been carried out
based on matter kinetic terms~\cite{BD1,BD2,BLR,KY,HKY,EI,Liu,DIU}.
}
From Eqs.(\ref{L-BSM-quark}) and (\ref{L-SM-quark-prime}),
the following matching conditions should be imposed on
\begin{eqnarray}
k_{ij}^{(q)} = \left\langle K_{ij}^{(q)} \right\rangle, ~~
k_{ij}^{(u)} = \left\langle K_{ij}^{(u)} \right\rangle,~~ 
k_{ij}^{(d)} = \left\langle K_{ij}^{(d)} \right\rangle,~~
\left(y_1\right)_{ij} = \left\langle \left(Y_1\right)_{ij} \right\rangle,~~
\left(y_2\right)_{ij} = \left\langle \left(Y_2\right)_{ij} \right\rangle,
\label{<KY>}
\end{eqnarray}
at $M_{\rm BSM}$.
From the fact that there are no exact flavor-dependent symmetries in the SM~\cite{LNS,Koide},
the common element of H$_{\rm F}^{\rm k}$ and H$_{\rm F}^{\rm y}$
should be a flavor-independent one.

We examine a relationship between 
the unitary bases ($q'_{\rm L}$, $u'_{\rm R}$, $d'_{\rm R}$)
and the SM quark fields denoted by 
non-prime ones ($q_{\rm L}$, $u_{\rm R}$, $d_{\rm R}$),
and study how flavor symmetries are realized in the SM ones. 
The unitary bases are, in general, related to 
the SM ones by the change of variables as
\begin{eqnarray}
&~& q_{\rm L} = N_q  q'_{\rm L},~~u_{\rm R} = N_u u'_{\rm R},~~d_{\rm R} = N_d d'_{\rm R},
\label{unitary-SM}
\end{eqnarray}
where $N_q$, $N_u$, and $N_d$ are $3 \times 3$ complex matrices
which are, in general, non-unitary matrices.
Under the transformation (\ref{F-tr-prime}),
the SM ones are transformed as
\begin{eqnarray}
q_{\rm L} \to \tilde{F}_{\rm L} q_{\rm L},~~
u_{\rm R} \to \tilde{F}_{\rm R}^{(u)} u_{\rm R},~~
d_{\rm R} \to \tilde{F}_{\rm R}^{(d)} d_{\rm R},~~
\phi \to e^{i \theta} \phi,
\label{tildeF-tr-prime}
\end{eqnarray}
where $\tilde{F}_{\rm L}$, $\tilde{F}_{\rm R}^{(u)}$, 
and $\tilde{F}_{\rm R}^{(d)}$ are defined by
\begin{eqnarray}
\tilde{F}_{\rm L} \equiv N_{q} F_{\rm L} N_{q}^{-1},~~
\tilde{F}_{\rm R}^{(u)} \equiv N_{u} F_{\rm R}^{(u)} N_{u}^{-1},~~
\tilde{F}_{\rm R}^{(d)} \equiv N_{d} F_{\rm R}^{(d)} N_{d}^{-1},
\label{tildeF-def}
\end{eqnarray}
respectively.
If $\tilde{F}_{\rm L}$, $\tilde{F}_{\rm R}^{(u)}$, 
and $\tilde{F}_{\rm R}^{(d)}$ belong to H$_{\rm F}^{\rm k}$,
they are unbroken elements realized by unitary matrices.
Otherwise, they are broken ones realized by non-unitary ones.
We call fields transformed by non-unitary matrices ``non-unitary bases''.

From the matching condition between the Lagrangian density (\ref{L-SM-quark-prime})
and that of the quark sector in the SM written by
\begin{eqnarray}
\mathscr{L}_{\rm SM}^{\rm quark} = \overline{q}_{{\rm L}i} i \SlashD q_{{\rm L}i}
+ \overline{u}_{{\rm R}i} i \SlashD u_{{\rm R}i}
+ \overline{d}_{{\rm R}i} i \SlashD d_{{\rm R}i}
 - y_{ij}^{(u)} \overline{q}_{{\rm L}i} \tilde{\phi} u_{{\rm R}j}
- y_{ij}^{(d)} \overline{q}_{{\rm L}i} \phi d_{{\rm R}j} + {\rm h.c.},
\label{L-Y-quark}
\end{eqnarray}
we obtain the relations:
\begin{eqnarray}
&~& k^{(q)}_{ij} = \left(N_{q}^{\dagger}N_{q}\right)_{ij},~~
k^{(u)}_{ij} = \left(N_{u}^{\dagger}N_{u}\right)_{ij},~~
k^{(d)}_{ij} = \left(N_{d}^{\dagger}N_{d}\right)_{ij},~~
\label{k-I}\\
&~& \left(y_1\right)_{ij} = \left(N_{q}^{\dagger} y^{(u)} N_{u}\right)_{ij},~~
\left(y_2\right)_{ij} = \left(N_{q}^{\dagger} y^{(d)} N_{d}\right)_{ij}.
\label{y-ySM}
\end{eqnarray}
Because the kinetic coefficients are hermitian and positive definite, 
$k_{ij}^{(q)}$ is written by
\begin{eqnarray}
k_{ij}^{(q)} = \left(U_q^{\dagger} (J_{q})^2 U_q\right)_{ij},
\label{kq-GF}
\end{eqnarray}
where $U_q$ is a $3 \times 3$ unitary matrix 
and $J_q$ is a real $3 \times 3$ diagonal matrix.
Then, $N_q$, $N_u$, and $N_d$ are parametrized by
\begin{eqnarray}
\hspace{-1cm}&~& N_q = V_q J_q U_q,~~ 
\label{Nq}\\
\hspace{-1cm}&~& N_u = \left(y^{(u)}\right)^{-1} \left(N_q^{\dagger}\right)^{-1} y_1
= \left(y^{(u)}\right)^{-1} V_q J_q^{-1} U_q y_1
= {V_{\rm R}^{(u)}}^{\dagger} \left(y_{\rm diag}^{(u)}\right)^{-1} V_{\rm L}^{(u)}
V_q J_q^{-1} U_q y_1,~~
\label{Nu}\\
\hspace{-1cm}&~& N_d = \left(y^{(d)}\right)^{-1} \left(N_q^{\dagger}\right)^{-1} y_2
= \left(y^{(d)}\right)^{-1} V_q J_q^{-1} U_q y_2
= {V_{\rm R}^{(d)}}^{\dagger} \left(y_{\rm diag}^{(d)}\right)^{-1} V_{\rm L}^{(d)}
V_q J_q^{-1} U_q y_2,
\label{Nd}
\end{eqnarray}
using $U_q$, $J_q$, a $3 \times 3$ unitary matrix $V_q$,
and the Yukawa coupling matrices $y^{(u)}$, $y^{(d)}$, $y_1$, and $y_2$.
In place of $y^{(u)}$ and $y^{(d)}$,
the diagonalized ones $y^{(u)}_{\rm diag}$ and $y^{(d)}_{\rm diag}$
and $3 \times 3$ unitary matrices
$V_{\rm L}^{(u)}$, $V_{\rm L}^{(d)}$, $V_{\rm R}^{(u)}$, and $V_{\rm R}^{(d)}$ are also used.
The $y^{(u)}$ and $y^{(d)}$ are diagonalized 
as $V_{\rm L}^{(u)} y^{(u)} {V_{\rm R}^{(u)}}^{\dagger} = y_{\rm diag}^{(u)}$
and $V_{\rm L}^{(d)} y^{(d)} {V_{\rm R}^{(d)}}^{\dagger} = y_{\rm diag}^{(d)}$,
and the quark masses are obtained as
\begin{eqnarray}
&~& V_{\rm L}^{(u)} y^{(u)} {V_{\rm R}^{(u)}}^{\dagger} \frac{v}{\sqrt{2}}
= y_{\rm diag}^{(u)} \frac{v}{\sqrt{2}} = M_{\rm diag}^{(u)} 
= {\rm diag}\left(m_u, m_c, m_t\right),
\label{Mu-diag}\\
&~& V_{\rm L}^{(d)} y^{(d)} {V_{\rm R}^{(d)}}^{\dagger} \frac{v}{\sqrt{2}}
= y_{\rm diag}^{(d)} \frac{v}{\sqrt{2}} = M_{\rm diag}^{(d)} 
= {\rm diag}\left(m_d, m_s, m_b\right),
\label{Md-diag}
\end{eqnarray}
where $v/\sqrt{2}$ is the VEV of the neutral component in the Higgs doublet, 
and $m_u$, $m_c$, $m_t$, $m_d$, $m_s$, and $m_b$ are masses of up, charm, top,
down, strange, and bottom quarks, respectively.
Using (\ref{Nu}) and (\ref{Nd}), $k_{ij}^{(u)}$ and $k_{ij}^{(d)}$ are rewritten by
\begin{eqnarray}
&~& k_{ij}^{(u)} = \left(y_1^{\dagger}{W^{(u)}}^{\dagger} 
\left(y_{\rm diag}^{(u) -1}\right)^2 W^{(u)} y_1\right)_{ij},
\label{ku-GF}\\
&~& k_{ij}^{(d)} = \left(y_2^{\dagger}{W^{(d)}}^{\dagger} 
\left(y_{\rm diag}^{(d) -1}\right)^2 W^{(d)} y_2\right)_{ij}
= \left(y_2^{\dagger}{W^{(u)}}^{\dagger} 
V_{\rm KM} \left(y_{\rm diag}^{(d) -1}\right)^2 
V_{\rm KM}^{\dagger} W^{(u)} y_2\right)_{ij},
\label{kd-GF}
\end{eqnarray}
where $W^{(u)} \equiv V_{\rm L}^{(u)} V_q J_{q}^{-1} U_q$,
$W^{(d)} \equiv V_{\rm L}^{(d)} V_q J_{q}^{-1} U_q$, 
and $V_{\rm KM} \equiv V_{\rm L}^{(u)} {V_{\rm L}^{(d)}}^{\dagger}$.
The $V_{\rm KM}$ is the Kobayashi$-$Maskawa matrix~\cite{KM}.
From Eq.(\ref{Nq}) and the definition of $W^{(u)}$,
we have the relations:
\begin{eqnarray}
N_q^{\dagger} = \left(W^{(u)}\right)^{-1} V_{\rm L}^{(u)},~~
N_q = {V_{\rm L}^{(u)}}^{\dagger} \left({W^{(u)}}^{\dagger}\right)^{-1}
\label{NqWu}
\end{eqnarray}
and, using them, we obtain the formula:
\begin{eqnarray}
k_{ij}^{(q)} = \left(W^{(u)}\right)^{-1}\left({W^{(u)}}^{\dagger}\right)^{-1}
= \left({W^{(u)}}^{\dagger}W^{(u)}\right)^{-1}.
\label{kq-GF2}
\end{eqnarray}
From the definition of $V_{\rm KM}$, 
we have the relation:
\begin{eqnarray}
W^{(u)} = V_{\rm KM} W^{(d)}~~~~{\rm or}~~~~W^{(d)} = V_{\rm KM}^{\dagger} W^{(u)}.
\label{Wud}
\end{eqnarray}
Note that $W^{(u)} $ and $W^{(d)}$ are not necessarily unitary matrices.
If $J_q$ is the identity matrix,
$k_{ij}^{(q)}$ is the canonical one ($\delta_{ij}$)
and $W^{(u)}$ and $W^{(d)}$ become unitary matrices.

\section{Chasing after flavor symmetries}

The $\mathscr{L}_{\rm SM}^{\rm quark}$ has been
obtained from accumulated experimental data and
successfully describes the physics of quarks at the weak scale.
Although the quark kinetic terms of $\mathscr{L}_{\rm SM}^{\rm quark}$
has the global ${\rm U}(3) \times {\rm U}(3) \times {\rm U}(3)/{\rm U}(1)$ symmetry,
this is an emergent symmetry and one takes care not to confuse it with 
flavor symmetries in an underlying theory described by
$\mathscr{L}_{\rm BSM}^{\rm quark}$.
Flavor symmetries are expected to be realized 
by unitary bases in ${\mathscr{L}'}_{\rm SM}^{\rm quark}$,
because it describes 
physics right after the change of flavor symmetries.
As ${\mathscr{L}'}_{\rm SM}^{\rm quark}$ 
can still retain the remnants of flavor symmetries
in spite of the fact that it is equivalent to $\mathscr{L}_{\rm SM}^{\rm quark}$,
it is favorable to examine ${\mathscr{L}'}_{\rm SM}^{\rm quark}$
in the pursuit of the origin of flavor.

\subsection{Generic argument}

We study generic properties of flavor symmetries 
based on ${\mathscr{L}'}_{\rm SM}^{\rm quark}$.
In Eqs.(\ref{ku-GF}) and (\ref{kd-GF}),
$k_{ij}^{(u)}$ and $k_{ij}^{(d)}$ are expanded as
\begin{eqnarray}
&~& k_{ij}^{(u)} = y_u^{-2} \left(y_1^{\dagger}{W^{(u)}}^{\dagger}\right)_{i1}
\left(W^{(u)} y_1\right)_{1j}
+ y_c^{-2} \left(y_1^{\dagger}{W^{(u)}}^{\dagger}\right)_{i2}
\left(W^{(u)} y_1\right)_{2j}
\nonumber \\
&~& ~~~~~~~~~~~~~ + y_t^{-2} \left(y_1^{\dagger}{W^{(u)}}^{\dagger}\right)_{i3}
\left(W^{(u)} y_1\right)_{3j},
\label{ku-GF-exp}\\
&~& k_{ij}^{(d)} = y_d^{-2} \left(y_2^{\dagger}{W^{(d)}}^{\dagger}\right)_{i1} 
\left(W^{(d)} y_2\right)_{1j}
+ y_s^{-2} \left(y_2^{\dagger}{W^{(d)}}^{\dagger}\right)_{i2} 
\left(W^{(d)} y_2\right)_{2j}
\nonumber \\
&~& ~~~~~~~~~~~~ + y_b^{-2} \left(y_2^{\dagger}{W^{(d)}}^{\dagger}\right)_{i3} 
\left(W^{(d)} y_2\right)_{3j},
\label{kd-GF-exp}
\end{eqnarray}
where $y_u$, $y_c$, $y_t$, $y_d$, $y_s$, and $y_b$ are
components of $y_{\rm diag}^{(u)}$ and $y_{\rm diag}^{(d)}$
and are estimated at the weak scale as
\begin{eqnarray}
&~& y_{\rm diag}^{(u)} = \left(y_u, y_c, y_t\right)
\doteqdot {\rm diag}\left(1.3 \times 10^{-5},~ 7.3 \times 10^{-3},~ 1.0\right),
\label{yu-diag-value}\\
&~& y_{\rm diag}^{(d)} = \left(y_d, y_s, y_b\right)
\doteqdot {\rm diag}\left(2.7 \times 10^{-5},~ 
5.5 \times 10^{-4},~ 2.4 \times 10^{-2}\right).
\label{yd-diag-value}
\end{eqnarray}
Physical parameters, in general, receive radiative corrections,
and the above values should be evaluated by considering 
renormalization effects to match with their counterparts at $M_{\rm BSM}$.

From the requirements that the magnitude of each component 
in $k_{ij}^{(u)}$ and $k_{ij}^{(d)}$ is at most $O(1)$
and there are no fine-tunings among terms including different couplings,
we obtain the conditions:
\begin{eqnarray}
&~& \left(W^{(u)} y_1\right)_{1j} \le O(y_u),~~
\left(W^{(u)} y_1\right)_{2j} \le O(y_c),~~
\left(W^{(u)} y_1\right)_{3j} \le O(y_t),~~
\label{Wuy1}\\
&~& \left(W^{(d)} y_2\right)_{1j} \le O(y_d),~~
\left(W^{(d)} y_2\right)_{2j} \le O(y_s),~~
\left(W^{(d)} y_2\right)_{3j} \le O(y_b).
\label{Wuy2}
\end{eqnarray}

Here, we explain some existence forms of flavor symmetries.
In an ordinary case,
fields belong to multiplets of irreducible representations of G$_{\rm F}$
and $\mathscr{L}_{\rm BSM}^{\rm quark}$ is constructed using G$_{\rm F}$-invariant
polynomials of fields.
There is a case that G$_{\rm F}$ appears as an accidental one from a more fundamental theory
and then fields can belong to multiplets of reducible representations effectively.
The ${\mathscr{L}'}_{\rm SM}^{\rm quark}$, in general, contains G$_{\rm F}$-invariant
and non-invariant parts of irreducible multiplets.
In some case, Yukawa interactions are composed of non-invariant terms alone.
In other case, an accidental flavor symmetry G$'_{\rm F}$ appears partially,
and ${\mathscr{L}'}_{\rm SM}^{\rm quark}$ contains invariant and non-invariant parts 
constructed from reducible multiplets.

As remnants of G$_{\rm F}$ in $\mathscr{L}_{\rm BSM}^{\rm quark}$
or an accidentalness of G$'_{\rm F}$ in ${\mathscr{L}'}_{\rm SM}^{\rm quark}$,
the kinetic coefficients $k_{ij}^{(x)}$ ($x=q, u, d$) 
and the Yukawa couplings $(y_1)_{ij}$ and $(y_2)_{ij}$, in general, consist of
flavor-symmetric and breaking parts
and are written as
\begin{eqnarray}
&~& k_{ij}^{(x)} = k_1^{(x)} \delta_{ij} 
+ k_{2}^{(x)} S_{ij}^{(x)} + \sum_{b_x} k_{3}^{(b_x)} T_{ij}^{(b_x)},
\label{k}\\
&~& (y_1)_{ij} = \sum_{a_1} y_{a_1}^{\rm F} S_{ij}^{(a_1)} + 
\sum_{b_1} \varDelta y_{b_1} T_{ij}^{(b_1)},~~
(y_2)_{ij} = \sum_{a_2} y_{a_2}^{\rm F} S_{ij}^{(a_2)} + 
\sum_{b_2} \varDelta y_{b_2} T_{ij}^{(b_2)}.
\label{y}
\end{eqnarray}
Here, terms containing $\delta_{ij}$ and $S_{ij}^{(A)}$ ($A = x, a_1, a_2$)
are flavor-symmetric parts,
(strictly speaking, flavor-dependent symmetric ones 
except for flavor-independent ones).
The $S_{ij}^{(A)}$ are $3 \times 3$ matrices (whose components take values of at most $O(1)$)
that satisfy the following relations from the G$_{\rm F}$ or G$'_{\rm F}$ invariance:
\begin{eqnarray}
&~& F_{\rm L}^{\dagger} S^{(q)} F_{\rm L} = S^{(q)},~~
F_{\rm R}^{(u)\dagger} S^{(u)} F_{\rm R}^{(u)} = S^{(u)},~~
F_{\rm R}^{(d)\dagger} S^{(d)} F_{\rm R}^{(d)} = S^{(d)},~~
\label{F-inv-Sk}\\
&~& e^{-i\theta} F_{\rm L}^{\dagger} S^{(a_1)} F_{\rm R}^{(u)} = S^{(a_1)},~~
e^{i\theta} F_{\rm L}^{\dagger} S^{(a_2)} F_{\rm R}^{(d)} = S^{(a_2)}.
\label{F-inv-Sy}
\end{eqnarray}
Note that several $S^{(a_1)}$s can exist, for example,
in the case that $q'_{\rm L}$ is a singlet but $u'_{\rm R}$
is a non-singlet of G$_{\rm F}$.
Terms containing $T_{ij}^{(B)}$ ($B = b_x, b_1, b_2$) 
are breaking ones.
The $T_{ij}^{(B)}$
are $3 \times 3$ matrices (whose components take values of at most $O(1)$).
For details, terms containing $T_{ij}^{(b_x)}$
are H$_{\rm F}^{\rm k}$ invariant ones
and those containing $T_{ij}^{(b_1)}$ and $T_{ij}^{(b_2)}$
are H$_{\rm F}^{\rm y}$ invariant ones.
In the absence of terms containing $T_{ij}^{(b_x)}$,
there should exist those containing $T_{ij}^{(b_1)}$ and $T_{ij}^{(b_2)}$
but no flavor symmetries must survive, 
from the fact that there are no exact flavor-dependent symmetries in the SM~\cite{LNS,Koide}

The coefficients $k_1^{(x)}$,  $k_2^{(x)}$, $k_3^{(b_x)}$, $y_{a_1}^{\rm F}$, 
$\varDelta y_{b_1}$, $y_{a_2}^{\rm F}$, and $\varDelta y_{b_2}$ are dimensionless parameters,
and the magnitude of their values can be a touchstone of new physics
by adopting Dirac's naturalness.
According to this concept,
we suppose that $k_1^{(x)}=O(1)$,  $k_2^{(x)}=(1)$, and
$\displaystyle{\left|y_{a_1}^{\rm F}\right| = O(1)}$ (for some $a_1$)
under the assumption that the relating terms originate from renormalizable interactions,
and, in contrast, magnitudes of other parameters can be tiny  
if their interactions stem from
non-renormalizable ones suppressed by a power of $M_{\rm BSM}$.
As a comment, some $\varDelta y_{b_1}$ and $\varDelta y_{b_2}$
can be sizable if the breaking scale of flavor symmetry is near $M_{\rm BSM}$.

In the following, we examine whether 
the magnitude of each component in $k_{ij}^{(u)}$ can be at most $O(1)$ or not, 
based on $\displaystyle{\left|y_{a_1}^{\rm F}\right| = O(1)}$ (for some $a_1$).

By inserting the first relation of Eq.(\ref{y}) into Eq.(\ref{ku-GF}),
we obtain the relation:
\begin{eqnarray}
&~& k_{ij}^{(u)} = \sum_{a_1, a'_1} y_{a_1}^{\rm F} y_{a'_1}^{{\rm F}*}
\left(S^{(a'_1)}{W^{(u)}}^{\dagger} 
\left(y_{\rm diag}^{(u) -1}\right)^2 W^{(u)} S^{(a_1)}\right)_{ij}
\nonumber \\
&~& ~~~~~~~~~~~~
+ \sum_{b_1, b'_1} \varDelta y_{b_1} \varDelta y_{b'_1}^*
\left({T^{(b'_1)}}^{\dagger}{W^{(u)}}^{\dagger} 
\left(y_{\rm diag}^{(u) -1}\right)^2 W^{(u)} T^{(b_1)}\right)_{ij}
\nonumber\\
&~& ~~~~~~~~~~~~ 
+ \sum_{a_1, b_1} y_{a_1}^{\rm F} \varDelta y_{b_1}^{*} 
\left({T^{(b_1)}}^{\dagger}{W^{(u)}}^{\dagger} 
\left(y_{\rm diag}^{(u) -1}\right)^2 W^{(u)} S^{(a_1)}\right)_{ij} + {\rm h.c.},
\label{ku-S}
\end{eqnarray}
and need the conditions:
\begin{eqnarray}
\hspace{-1.2cm}&~& y_{a_1}^{\rm F} \left(W^{(u)} S^{(a_1)}\right)_{1j} \le O(y_u),~~
y_{a_1}^{\rm F} \left(W^{(u)} S^{(a_1)}\right)_{2j} \le O(y_c),~~
y_{a_1}^{\rm F} \left(W^{(u)} S^{(a_1)}\right)_{3j} \le O(y_t),~~
\label{WuS1}\\
\hspace{-1.2cm}&~& \varDelta y_{b_1} \left(W^{(u)} T^{(b_1)}\right)_{1j} \le O(y_u),~~
\varDelta y_{b_1} \left(W^{(u)} T^{(b_1)}\right)_{2j} \le O(y_c),~~
\varDelta y_{b_1} \left(W^{(u)} T^{(b_1)}\right)_{3j} \le O(y_t),~~
\label{WuT1}
\end{eqnarray}
in order to make the magnitudes of $k_{ij}^{(u)}$ at most $O(1)$,
unless any cancellations occur among several contributions.
If the magnitude of $\displaystyle{\left(W^{(u)} T^{(b_1)}\right)_{ij}}$ is $O(1)$,
the conditions (\ref{WuT1}) fulfill 
with $\displaystyle{\left|\varDelta y_{b_1}\right| = O(y_u)}$.
In the case that 
the magnitude of $\displaystyle{\left(W^{(u)} T^{(b_1)}\right)_{1j}}$ 
is $O(y_u/y_c)$,
that of $\displaystyle{\left|\varDelta y_{b_1}\right|}$ can be $O(y_c)$.
Furthermore, in the case that 
the magnitude of $\displaystyle{\left(W^{(u)} T^{(b_1)}\right)_{1j}}$ 
and $\displaystyle{\left(W^{(u)} T^{(b_1)}\right)_{2j}}$ 
are $O(y_u/y_t)$ and $O(y_c/y_t)$, respectively,
that of $\displaystyle{\left|\varDelta y_{b_1}\right|}$ can be $O(y_t)$.
This suggests that a mass hierarchy of up-type quarks
can be realized by the breaking part alone.

Hereafter, we consider a case with $\displaystyle{\left|y_{a_1}^{\rm F}\right| = O(1)}$
and $\displaystyle{\left(W^{(u)} S^{(a_1)}\right)_{3j} = O(1)}$ (for some $a_1$)
under the assumption that 
$y_{a_1}^{\rm F}S^{(a_1)}$ comes from a renormalizable interaction.
Then, we find that $\displaystyle{S_{ij}^{(a_1)} = \left({W^{(u)}}^{-1}\right)_{i3} 
\left(W^{(u)} S^{(a_1)}\right)_{3j}}$
up to $O(y_c)$, in the case that the magnitude of each component of $W^{(u)}$ is $O(1)$, 
from the conditions (\ref{WuS1}).
In most cases, tiny quantities of $O(y_c)$
and $O(y_u)$ can appear from symmetry breaking effects,
and hence we suppose that 
$\displaystyle{S_{ij}^{(a_1)} = \left({W^{(u)}}^{-1}\right)_{i3} 
\left(W^{(u)} S^{(a_1)}\right)_{3j}}$
holds exactly in a flavor-symmetric limit.
In this case, after a suitable unitary transformation is performed,
${S^{(a_1)}}^{\dagger} S^{(a_1)}$ is diagonalized as
\begin{eqnarray}
U \left({S^{(a_1)}}^{\dagger} S^{(a_1)}\right) U^{\dagger} = 
\left(
\begin{array}{ccc}
0 & 0 & 0 \\
0 & 0 & 0 \\
0 & 0 & s
\end{array} 
\right),
\label{S1S1}
\end{eqnarray}
where $s$ is given by
\begin{eqnarray}
s = \sum_{k=1}^3 \left|\left({W^{(u)}}^{-1}\right)_{k3}\right|^2
\sum_{l=1}^3 \left|\left(W^{(u)} S^{(a_1)}\right)_{3l}\right|^2.
\label{s}
\end{eqnarray}
This implies that $S^{(a_1)}$ is a $3 \times 3$ matrix whose rank is one.

In the same way, 
by inserting the second relation of Eq.(\ref{y}) into Eq.(\ref{kd-GF}),
we obtain the relation:
\begin{eqnarray}
&~& k_{ij}^{(d)} = \sum_{a_2, a'_2} y_{a_2}^{\rm F} y_{a'_2}^{{\rm F}*}
\left(S^{(a'_2)}{W^{(d)}}^{\dagger} 
\left(y_{\rm diag}^{(d) -1}\right)^2 W^{(d)} S^{(a_2)}\right)_{ij}
\nonumber \\
&~& ~~~~~~~~~~~~
+ \sum_{b_2, b'_2} \varDelta y_{b_2} \varDelta y_{b'_2}^*
\left({T^{(b'_2)}}^{\dagger}{W^{(d)}}^{\dagger} 
\left(y_{\rm diag}^{(d) -1}\right)^2 W^{(d)} T^{(b_2)}\right)_{ij}
\nonumber\\
&~& ~~~~~~~~~~~~ 
+ \sum_{a_2, b_2} y_{a_2}^{\rm F} \varDelta y_{b_2}^{*} 
\left({T^{(b_2)}}^{\dagger}{W^{(d)}}^{\dagger} 
\left(y_{\rm diag}^{(d) -1}\right)^2 W^{(d)} S^{(a_2)}\right)_{ij} + {\rm h.c.},
\label{kd-S}
\end{eqnarray}
and need the conditions:
\begin{eqnarray}
\hspace{-1.2cm}&~& y_{a_2}^{\rm F} \left(W^{(d)} S^{(a_2)}\right)_{1j} \le O(y_d),~~
y_{a_2}^{\rm F} \left(W^{(d)} S^{(a_2)}\right)_{2j} \le O(y_s),~~
y_{a_2}^{\rm F} \left(W^{(d)} S^{(a_2)}\right)_{3j} \le O(y_b),~~
\label{WdS2}\\
\hspace{-1.2cm}&~& \varDelta y_{b_2} \left(W^{(d)} T^{(b_2)}\right)_{1j} \le O(y_d),~~
\varDelta y_{b_2} \left(W^{(d)} T^{(b_2)}\right)_{2j} \le O(y_s),~~
\varDelta y_{b_2} \left(W^{(d)} T^{(b_2)}\right)_{3j} \le O(y_b),~~
\label{WdT2}
\end{eqnarray}
in order to make the magnitudes of $k_{ij}^{(d)}$ at most $O(1)$,
unless any cancellations occur among several contributions.
If the magnitude of $\displaystyle{\left(W^{(d)} T^{(b_2)}\right)_{1j}}$ is $O(1)$,
the conditions (\ref{WdT2}) fulfill 
with $\displaystyle{\left|\varDelta y_{b_2}\right| = O(y_d)}$.
In the case that 
the magnitude of $\displaystyle{\left(W^{(d)} T^{(b_2)}\right)_{1j}}$ is $O(y_d/y_s)$,
that of $\displaystyle{\left|\varDelta y_{b_2}\right|}$ can be $O(y_s)$.
Furthermore, in the case that 
the magnitude of $\displaystyle{\left(W^{(d)} T^{(b_2)}\right)_{1j}}$ 
and $\displaystyle{\left(W^{(d)} T^{(b_2)}\right)_{2j}}$ 
are $O(y_d/y_b)$ and $O(y_s/y_b)$, respectively,
that of $\displaystyle{\left|\varDelta y_{b_2}\right|}$ can be $O(y_b)$.
This also suggests that a mass hierarchy of down-type quarks
can be realized by the breaking part alone.

From (\ref{WdS2}), it is conjectured that 
that $y_{a_2}^{\rm F}S^{(a_2)}$ can also stem from non-renormalizable interactions,
i.e., $\displaystyle{\left|y_{a_2}^{\rm F}\right| \le O(y_b)}$, 
if $\displaystyle{\left(W^{(d)} S^{(a_2)}\right)_{3j} = O(1)}$.
For instance, a down-type quark Yukawa coupling matrix can be
obtained by the Froggatt-Nielsen mechanism of a flavor-independent charge
with $F_{\rm L} = e^{i\varphi_{\rm L}}I$ and $F_{\rm R}^{(d)} = e^{i\varphi_{\rm R}^{(d)}}I$
from a non-renormalizable term 
$\left(Y_2\right)_{ij} \overline{q}'_{{\rm L}i} \phi d'_{{\rm R}j}$
where $\left(Y_2\right)_{ij}$ contains 
$\displaystyle{\left({\varphi}/{\varLambda}\right)^n}$~\cite{FN}.
Here, $\varphi$ is the SM-singlet scalar field with the VEV of $O(M_{\rm BSM})$
and $\varLambda$ is a cutoff scale bigger than $M_{\rm BSM}$.
If the magnitude of $\left(W^{(d)} S^{(a_2)}\right)_{3j}$
is much bigger than that of $\left(W^{(d)} S^{(a_2)}\right)_{1j}$ 
and $\left(W^{(d)} S^{(a_2)}\right)_{2j}$
and the magnitude of each component of $W^{(d)}$ is $O(1)$,
$\displaystyle{S_{ij}^{(a_2)} = \left({W^{(d)}}^{-1}\right)_{i3} 
\left(W^{(d)} S^{(a_2)}\right)_{3j}}$
up to $O(y_{s})$.
In the case that $\displaystyle{S_{ij}^{(a_2)} 
= \left({W^{(d)}}^{-1}\right)_{i3} \left(W^{(d)} S^{(a_2)}\right)_{3j}}$
holds exactly,
${S^{(a_2)}}^{\dagger} S^{(a_2)}$ is also diagonalized as
the same form of (\ref{S1S1}) with 
\begin{eqnarray}
s = \sum_{k=1}^3 \left|\left({W^{(d)}}^{-1}\right)_{k3}\right|^2
\sum_{l=1}^3 \left|\left(W^{(d)} S^{(a_2)}\right)_{3l}\right|^2,
\label{s2}
\end{eqnarray}
and $S^{(a_2)}$ is also a $3 \times 3$ matrix whose rank is one.

Under the assumption that $\displaystyle{\left|y_{a_1}^{\rm F}\right| = O(1)}$,
$\displaystyle{\left(W^{(u)} S^{(a_1)}\right)_{3j} = O(1)}$
and $S^{(a_1)} = S^{(a_2)}$,
the magnitude of $\displaystyle{\left|y_{a_2}^{\rm F}\right|}$ is estimated as follows.
Using Eq.(\ref{Wud}), we obtain the relation:
\begin{eqnarray}
y_{a_2}^{\rm F} W^{(d)} S^{(a_2)}
= y_{a_2}^{\rm F} V_{\rm KM}^{\dagger} W^{(u)} S^{(a_1)}
= y_{a_2}^{\rm F}
\left(
\begin{array}{ccc}
O(\lambda^3) & O(\lambda^3) & O(\lambda^3) \\
O(\lambda^2) & O(\lambda^2) & O(\lambda^2) \\
O(1) & O(1) & O(1)
\end{array} 
\right),
\label{WdS2-2}
\end{eqnarray}
where $\lambda = \sin\theta_{\rm C} \doteqdot 0.225$ 
($\theta_{\rm C}$ is the Cabibbo angle~\cite{C}),
and we use the Wolfenstein parametrization~\cite{Wolf}.
From the conditions (\ref{WdS2}) and Eq.(\ref{WdS2-2}),
we derive the inequality:
\begin{eqnarray}
\left|y_{a_2}^{\rm F}\right| \le O\left(y_d/\lambda^3\right)
= O(10^{-3}).
\label{ya2F}
\end{eqnarray}

In this way, we have obtained the following properties.
\begin{itemize}
\item The magnitude of $\displaystyle{y_{a_1}^{\rm F}}$
can be $O(1)$ and some $y_{a_1}^{\rm F} S_{ij}^{(a_1)}$ can appear from
a renormalizable interaction in a theory beyond the SM.

\item The magnitudes of $\displaystyle{y_{a_2}^{\rm F}}$
can be $O(y_b)=O(10^{-2})$ or less than that,
and $y_{a_2}^{\rm F} S_{ij}^{(a_2)}$ can appear from
non-renormalizable interactions 
through the Froggatt-Nielsen mechanism.

\item Some $S_{ij}^{(a_1)}$ and $S_{ij}^{(a_2)}$ can be rank-one matrices.

\item The magnitude of $\displaystyle{y_{a_2}^{\rm F}}$ can be 
$O\left(y_d/\lambda^3\right)=O(10^{-3})$ or less than that,
in the case with $\displaystyle{\left|y_{a_1}^{\rm F}\right|=O(1)}$
and $S_{ij}^{(a_1)} = S_{ij}^{(a_2)}$.
\end{itemize}

\subsection{Peculiarity of democratic type}

If $N_q$, $N_u$, and $N_d$ are given, $k_{ij}^{(q)}$, $k_{ij}^{(u)}$, $k_{ij}^{(d)}$, 
$\left(y_1\right)_{ij}$, and $\left(y_2\right)_{ij}$ are determined
by Eqs.(\ref{k-I}) and (\ref{y-ySM}).
If $W^{(u)}$, $\left(y_1\right)_{ij}$, and $\left(y_2\right)_{ij}$ are given,
$k_{ij}^{(q)}$, $k_{ij}^{(u)}$, and $k_{ij}^{(d)}$ are determined
by Eqs.(\ref{kq-GF2}), (\ref{ku-GF}), and (\ref{kd-GF}).
In the following, we show that 
the flavor-symmetric parts of  
$\left(y_1\right)_{ij}$ and $\left(y_2\right)_{ij}$
and parts of $k_{ij}^{(u)}$ and $k_{ij}^{(d)}$
constructed from them are constrained
and a democratic-type matrix takes a special position, supposing that $W^{(u)}$ is given
and Dirac's naturalness is adopted.

Let the Yukawa couplings be divided into two parts as
\begin{eqnarray}
\left(y_1\right)_{ij} = \left(y_1^{\rm F}\right)_{ij} 
+ \left(y_1^{\SlashF}\right)_{ij},~~
\left(y_2\right)_{ij} = \left(y_2^{\rm F}\right)_{ij} 
+ \left(y_2^{\SlashF}\right)_{ij},
\label{y-FslashF}
\end{eqnarray}
where $\displaystyle{\left(y_1^{\rm F}\right)_{ij}}$ 
and $\displaystyle{\left(y_2^{\rm F}\right)_{ij}}$ are flavor-symmetric parts,
and $\displaystyle{\left(y_1^{\SlashF}\right)_{ij}}$ 
and $\displaystyle{\left(y_2^{\SlashF}\right)_{ij}}$ are flavor-breaking ones.
Using $\displaystyle{\left(y_1^{\rm F}\right)_{ij}}$ 
and $\displaystyle{\left(y_2^{\rm F}\right)_{ij}}$,
we define $\displaystyle{\tilde{k}_{ij}^{(u)}}$ 
and $\displaystyle{\tilde{k}_{ij}^{(d)}}$ as
\begin{eqnarray}
&~& \tilde{k}_{ij}^{(u)} \equiv \left({y_1^{\rm F}}^{\dagger}{W^{(u)}}^{\dagger} 
\left(y_{\rm diag}^{(u) -1}\right)^2 W^{(u)} y_1^{\rm F}\right)_{ij},
\label{ku-GF-F}\\
&~& \tilde{k}_{ij}^{(d)} \equiv \left({y_2^{\rm F}}^{\dagger}{W^{(d)}}^{\dagger} 
\left(y_{\rm diag}^{(d) -1}\right)^2 W^{(d)} y_2^{\rm F}\right)_{ij}.
\label{kd-GF-F}
\end{eqnarray}
In the case that $k_{ij}^{(q)}$ is flavor symmetric, i.e.,
$F_{\rm L} k_{ij}^{(q)} F_{\rm L}^{\dagger} = k_{ij}^{(q)}$,
or in a flavor-symmetric limit (after neglecting the breaking parts in $k_{ij}^{(q)}$),
$\displaystyle{\tilde{k}_{ij}^{(u)}}$ 
and $\displaystyle{\tilde{k}_{ij}^{(d)}}$ also become flavor symmetric.

From the requirements that the magnitude of each component 
in $k_{ij}^{(u)}$ is at most $O(1)$, 
$\displaystyle{\tilde{k}_{ij}^{(u)}}$ contains a parameter of $O(1)$
such as $y_t$
and any tiny parameters 
are not included,
the form of $\displaystyle{\left(y_1^{\rm F}\right)_{ij}}$ is constrained as
\begin{eqnarray}
y_1^{\rm F} = 
\left(
\begin{array}{ccc}
l v_1 & m v_1 & n v_1 \\
l v_2 & m v_2 & n v_2 \\
l v_3 & m v_3 & n v_3 
\end{array} 
\right),
\label{y1-F}
\end{eqnarray}
where $l$, $m$, and $n$ are some numbers, and
$v_1$, $v_2$, and $v_3$ are defined by
\begin{eqnarray}
&~& v_1 \equiv W_{12}^{(u)} W_{23}^{(u)} - W_{13}^{(u)} W_{22}^{(u)},~~
v_2 \equiv W_{13}^{(u)} W_{21}^{(u)} - W_{11}^{(u)} W_{23}^{(u)},~~
\nonumber \\
&~& v_3 \equiv W_{11}^{(u)} W_{22}^{(u)} - W_{12}^{(u)} W_{21}^{(u)}.
\label{v123}
\end{eqnarray}
Eqs.(\ref{v123}) are derived from the orthogonality
between $\displaystyle{\left(W_{11}^{(u)},W_{12}^{(u)},W_{13}^{(u)}\right)}$ and $(v_1, v_2, v_3)$,
and $\displaystyle{\left(W_{21}^{(u)},W_{22}^{(u)},W_{23}^{(u)}\right)}$ and $(v_1, v_2, v_3)$.
Then, $\displaystyle{\tilde{k}_{ij}^{(u)}}$ is written by
\begin{eqnarray}
\tilde{k}^{(u)} = \frac{1}{y_t^2}
\left(W_{31}^{(u)} v_1 + W_{32}^{(u)} v_2 + W_{33}^{(u)} v_3\right)^2
\left(
\begin{array}{ccc}
l^2 & m l & n l \\
l m & m^2 & n m \\
l n & m n & n^2 
\end{array} 
\right).
\label{ku-GF-F2}
\end{eqnarray}

Next, we attempt to conjecture a flavor symmetry by imposing on
$F_{\rm R}^{(u)} \tilde{k}^{(u)} {F_{\rm R}^{(u)}}^{\dagger} = \tilde{k}^{(u)}$.
In the case with $l = m = n$,
$\tilde{k}^{(u)}$ becomes a democratic-type matrix, which is proportional to the matrix:
\begin{eqnarray}
S = 
\left(
\begin{array}{ccc}
1 & 1 & 1 \\
1 & 1 & 1 \\
1 & 1 & 1
\end{array} 
\right),
\label{democratic}
\end{eqnarray}
and $\displaystyle{\left(y_1^{\rm F}\right)_{ij}}$ also 
turns out to be the democratic-type one
for $l = m = n$ and $v_1 = v_2 = v_3$.
This form has an invariance 
under a discrete group such as S$_3$,
where fields are transformed as a 3D reducible representation.
\footnote{
Based on an ${\rm S}_3$ invariant K\"{a}hler potential
containing the democratic form and Yukawa couplings
with the democratic form and small S$_3$ breaking ones,
it was pointed out that the heavy top quark mass can be attributed to
a singular normalization of its kinetic term~\cite{KY}.
}
Actually, the permutations of reducible triplet
are performed by the matrices:
\begin{eqnarray}
&~& U^{\alpha} = \left(
\begin{array}{ccc}
1 & 0 & 0 \\
0 & 1 & 0 \\
0 & 0 & 1
\end{array} 
\right),~~ 
\left(
\begin{array}{ccc}
0 & 1 & 0 \\
1 & 0 & 0 \\
0 & 0 & 1
\end{array} 
\right),~~
\left(
\begin{array}{ccc}
1 & 0 & 0 \\
0 & 0 & 1 \\
0 & 1 & 0
\end{array} 
\right),~~
\left(
\begin{array}{ccc}
0 & 0 & 1 \\
0 & 1 & 0 \\
1 & 0 & 0
\end{array} 
\right),~~
\nonumber \\
&~& ~~~~~~~~~~~ \left(
\begin{array}{ccc}
0 & 1 & 0 \\
0 & 0 & 1 \\
1 & 0 & 0
\end{array} 
\right),~~
\left(
\begin{array}{ccc}
0 & 0 & 1 \\
1 & 0 & 0 \\
0 & 1 & 0
\end{array} 
\right),
\label{S3}
\end{eqnarray}
and $S$ is constructed as
\begin{eqnarray}
S = \frac{1}{2} \sum_{\alpha=1}^{6} U^{\alpha}.
\label{Sa1}
\end{eqnarray}
The invariance is understood from
the relations for any elements $U^{\beta}$ ($\beta = 1, \cdots, 6$):
\begin{eqnarray}
\left(\sum_{\alpha=1}^{6} U^{\alpha}\right) U^{\beta} 
= U^{\beta} \left(\sum_{\alpha=1}^{6} U^{\alpha}\right)  
= \sum_{\alpha=1}^{6} U^{\alpha},
\label{Sa1-inv}
\end{eqnarray}
where $U^{\alpha} U^{\beta} \ne U^{\alpha'} U^{\beta}$
and $U^{\beta} U^{\alpha} \ne U^{\beta} U^{\alpha'}$
for $U^{\alpha} \ne U^{\alpha'}$.

In the case with $l = m$ and $n \ne l$,
$\tilde{k}^{(u)}$ is proportional to the matrix:
\begin{eqnarray}
S = 
\left(
\begin{array}{ccc}
1 & 1 & n \\
1 & 1 & n \\
n & n & n^2/l
\end{array} 
\right),
\label{semi-democratic}
\end{eqnarray}
and this form has an invariance under a discrete group such as S$_2$.
For $l \ne m$, $m \ne n$ and $n \ne l$,
$F_{\rm R}^{(u)}$ is proportional to the identity matrix,
and there is no flavor-dependent symmetry in $\tilde{k}^{(u)}$.

In the same way, from the requirements that the magnitude of each component 
in $k_{ij}^{(d)}$ is at most $O(1)$ and any tiny parameters 
except for $y_b$ are not included in $\displaystyle{\tilde{k}_{ij}^{(d)}}$,
the form of $\displaystyle{\left(y_2^{\rm F}\right)_{ij}}$ is constrained as
\begin{eqnarray}
y_2^{\rm F} = 
\left(
\begin{array}{ccc}
p w_1 & q w_1 & r w_1 \\
p w_2 & q w_2 & r w_2 \\
p w_3 & q w_3 & r w_3 
\end{array} 
\right),
\label{y2-F}
\end{eqnarray}
where $p$, $q$, and $r$ are some numbers, and
$w_1$, $w_2$, and $w_3$ are defined by
\begin{eqnarray}
&~& w_1 \equiv W_{12}^{(d)} W_{23}^{(d)} - W_{13}^{(d)} W_{22}^{(d)},~~
w_2 \equiv W_{13}^{(d)} W_{21}^{(d)} - W_{11}^{(d)} W_{23}^{(d)},~~
\nonumber \\
&~& w_3 \equiv W_{11}^{(d)} W_{22}^{(d)} - W_{12}^{(d)} W_{21}^{(d)}.
\label{w123}
\end{eqnarray}
Then, $\displaystyle{\tilde{k}_{ij}^{(d)}}$ is written by
\begin{eqnarray}
\tilde{k}^{(d)} = \frac{1}{y_b^2}
\left(W_{31}^{(d)} w_1 + W_{32}^{(d)} w_2 + W_{33}^{(d)} w_3\right)^2
\left(
\begin{array}{ccc}
p^2 & q p & r p \\
p q & q^2 & r q \\
p r & q r & r^2 
\end{array} 
\right).
\label{kd-GF-F2}
\end{eqnarray}
Note that $|p|, |q|, |r| \le O(y_b)$
in order to make the magnitude of $\tilde{k}_{ij}^{(d)}$ at most $O(1)$.
We consider a flavor symmetry on down-type quarks.
For $p = q = r$, $\tilde{k}^{(d)}$ becomes the democratic one.
For $v_1 = v_2 = v_3$, $w_1 = w_2 = w_3$ does not hold
because of Eq.(\ref{Wud})
and then $\displaystyle{\left(y_2^{\rm F}\right)_{ij}}$
cannot be a democratic one.

Finally, we give a comment on a case that $\displaystyle{\left(y_2^{\rm F}\right)_{ij}}$
is a democratic one $\displaystyle{\left(y_2^{\rm F}\right)_{ij}
=\tilde{y}_2^{\rm F}S_{ij}}$ with a complex number $\tilde{y}_2^{\rm F}$.
In this case, $\tilde{k}^{(d)}$ also becomes the democratic one:
\begin{eqnarray}
&~& \tilde{k}_{ij}^{(d)} = \left|\tilde{y}_2^{\rm F}\right|^2
\left(S{W^{(u)}}^{\dagger} V_{\rm KM}
\left(y_{\rm diag}^{(d) -1}\right)^2 V_{\rm KM}^{\dagger}
W^{(u)}S\right)_{ij}
\nonumber \\
&~& ~~~~~~~~~
= O\left(\lambda^6/y_d^{2}\right) \left|\tilde{y}_2^{\rm F}\right|^2
\left(W_{31}^{(u)} + W_{32}^{(u)} + W_{33}^{(u)}\right)^2 S_{ij}.
\label{kd-GF-S}
\end{eqnarray}
The following inequality is required
\begin{eqnarray}
\left|\tilde{y}_{2}^{\rm F}\right| \le O\left(y_d/\lambda^3\right)
= O\left(10^{-3}\right)
\label{y2F}
\end{eqnarray}
to make the magnitude of each component in $\tilde{k}_{ij}^{(d)}$ at most $O(1)$.

In this way, we find that the democratic-type one takes a special position,
because it is related to a flavor symmetry such as S$_3$
and is compatible with Dirac's naturalness.

\section{Conclusions and discussions}

We have explored the flavor structure in the SM
under the assumption that flavor symmetries exist 
in a theory beyond the SM,
and have chased after their properties, using a bottom-up approach.
We have reacknowledged that a flavor-symmetric part 
of Yukawa coupling matrix can be realized by a rank-one matrix
and a democratic-type one occupies a special position, based on Dirac's naturalness.
Hence, it would be important to explore the origin of the democratic-type matrix.
There is a possibility that it is generated by the VEVs of flavons.
However, a toy model presented in \cite{YK} has a problem 
that it contains an unnatural fine-tuning among parameters
based on a perturbative analysis.
A non-perturbative effect can play a crucial role to the derivation of
a specific type of terms.

There are limitations on our bottom-up approach,
without any powerful principle and concept.
It would be desirable to combine use of the bottom-up
and top-down ones, 
keeping an eye on the possibility of grand unification and supersymmetry (SUSY).
On a grand unification based on $SO(10)$ and $E_6$,
we need an extension of Yukawa sector.
Without extra matters and/or extra interactions,
it is difficult to derive realistic fermion masses and flavor mixing matrices
in the case that a flavor-symmetric part dominates.
The reason is as follows.
Both $u'_{\rm R}$ and $d'_{\rm R}$ belong to a common multiplet
of $SO(10)$ and $E_6$, 
they should be transformed as a same representation of same flavor group,
and their kinetic coefficients have a common one, i.e., $S^{(a_1)} = S^{(a_2)}$.
Then, a common Yukawa coupling constant of $O(1)$
is not compatible with $S^{(a_1)} = S^{(a_2)}$.
The SUSY can compensate for the lack of information on the flavor structure,
that is, a pattern of soft SUSY breaking terms can provide
useful information.
It would be worth studying the flavor structure of the SM and its underlying theory
by paying close attention to both matter kinetic terms and various interaction terms.

\section*{Acknowledgments}
This work was supported in part by scientific grants 
from the Ministry of Education, Culture,
Sports, Science and Technology under Grant No.~17K05413.

\end{document}